\documentstyle[12pt]{article}

\begin{document}

\title{UNITARITY BOUNDS IN THE VECTOR CONDENSATE  
MODEL OF ELECTROWEAK INTERACTIONS}

\author{ G. Cynolter, A. Bodor and G. P\'ocsik \\
  Institute for Theoretical Physics, \\
  E\"otv\"os Lorand University, Budapest }
\date{ITP-Budapest Report No. 515. \\ February 1996} 
\maketitle 
 
\begin{abstract} 
 
{We replace the standard model scalar doublet by a  
doublet of vector fields  
and generate masses by dynamical symmetry breaking. 
Oblique radiative corrections are small if the  
new vector bosons ($ B^+$, $ B^0 $) are heavy. 
In this note it is shown that the model has a  
low momentum scale and above  
$ \Lambda \simeq 2 $ TeV it does not respect the  
perturbative unitarity. From tree-graph unitarity the allowed  
region of $B ^+ (B^0) $ mass is estimated as   
$m_{B^+} \geq 369 \hbox{GeV} \quad (m_{B^0} \geq 410 $ GeV) 
at $\Lambda= 1 $ TeV.} 
 
\end{abstract} 
  
Motivated by the fact that the Higgs is not yet seen, recently, 
we have proposed a version of the standard model [1,2] making 
use of dynamical symmetry breaking, where the scalar doublet is 
replaced by a vector doublet, 
\begin{equation} 
 B_{ \mu } = \pmatrix{ B_{ \mu }^{(+)} \cr 
 B_{ \mu }^{(0)} \cr},
\end{equation}
and its neutral member forms a nonvanishing condensate 
\begin{equation} 
 \left \langle B_{ \mu }^{(0)+} B_{ \nu }^{(0)} \right \rangle_0  
  = g_{ \mu \nu } d , \qquad d \not = 0 . 
\end{equation}	 
Gauge boson and $B^{+,0} $ boson masses are coming 
from the gauge invariant Lagrangian 
\begin{equation} 
L=-{1 \over 2} \left( D_{\mu}B_{\nu}-D_{\nu}B_{\mu} 
\right)^+ \left( D^{\mu}B^{\nu}-D^{\nu}B^{\mu} \right) 
-\lambda (B_{\nu}^+ B^{\nu} )^2, \lambda >0 
\end{equation} 
with $D_{\mu}$ the usual covariant derivative in the standard model. 
Fermion masses are also generated by (2) [2], as well as 
fermion and gauge field couplings are standard. 
Ratio of bare $B^{+,0} $ masses is ${ (4/5)}^{\frac{1}{2}}$ . 
From low energy charged current  
phenomenology we find  
$\sqrt{-6d}=(\sqrt{2} G_F )^{-\frac{1}{2}}=246 $ GeV. \linebreak
$ \lambda $ is proportional to the bare $m_0^2$ [1], 
$ m_0^2=-10 d \lambda$. 
 
It has been shown that oblique radiative  
corrections can be suppressed by  
choosing heavy $ B^{+,0} $ bosons [2,3]. For instance, at a scale of 
$ \Lambda=1 $ TeV, for $ S_{new} \geq 0 $ 
the thresholds for physical masses are about 
$m_0 \simeq 400 $ GeV, $ m_+ \simeq 200 $ GeV.  
Furthermore, higher $\Lambda$   allows higher minimum masses, 
but $\Lambda$  remains  
unrestricted even if the condensate (2) makes  
it plausible that the momentum scale is a few TeV's.  
\bigskip
In the present note we apply tree--level partial--wave 
unitarity to two-body scatterings of longitudinal gauge  
and B bosons following the reasoning in ref. [4]  
where perturbative unitarity has been employed for  
constraining the Higgs mass.  
We get that for scales $\Lambda \geq 2 $ TeV 
partial-wave unitarity breaks down, and B  
masses are bounded from below. 
The lower bound is increasing with growing $\Lambda$.
At $\Lambda=1 $ TeV the best bounds we find are 
 $m_+ \geq 369 $ GeV $ m_0 \geq 410 $ GeV. 
We remark that B--particles could be copiously produced  
at high-energy linear $e^+e^-$  
colliders up to masses of several hundred GeV's [5]. 

In the vector condensate model there exist many 
$ BB \rightarrow BB,VV $; $ BV \rightarrow BV $  
type processes with 
$B=B^0, \overline{B}^0, B^{\pm}$, $ V=W^{\pm}, Z$. 
We consider them for longitudinally  
polarized external particles and calculate the  
J=0 partial-wave amplitudes, $a_0$,  
from contact and one-particle exchange graphs.  
Unitarity requires $ \vert Re a_0 \vert \leq \frac{1}{2} $. 
 
The trilinear interactions of $Z, W^{\pm} $ 
are derived from (3) as 
 
\begin{eqnarray} 
 L  \left( B^{0} \right)&=&{ig \over 2cos \Theta_w } \partial^{\mu}
 B^{(0)\nu +} \left( Z_{ \mu} B_{\nu}^{(0)} - Z_{\nu} B^{(0)}_{\mu}
\right) + h.c., \nonumber \\ 
 L \left( B^+ {B^-} Z \right)&=&-(cos2\Theta_w) \cdot L
\left(B^{(0)} \rightarrow B^{(+)} \right), \\ 
 L  \left( B^{0}B^+W \right)&=&{ig \over \sqrt{2}} \Big[ \partial^{\mu} 
 B^{(+)\nu +} \left( W_{ \mu}^+ B_{\nu}^{(0)} -W_{\nu}^+
B^{(0)}_{\mu}\right) + \nonumber \\
& & + \partial^{\mu}  B^{(0)\nu +}  
\left( W_{ \mu}^- B_{\nu}^{(+)} -W_{\nu}^- B^{(+)}_{\mu}
\right) \Big] + h.c. \nonumber
\end{eqnarray}

The quartic interactions coming from (3) are self-couplings
of $ B^{+,0} $ and couplings of the type
$ \gamma \gamma B^+B^-$, $ZZ B^+ B^- $, $\gamma ZB^+ B^-$,
$\gamma W^+B^-B^0 $, $ZW^+B^-B^0 $,  $W^-W^+B^-B^+ $, 
$ W^+ W^- B^0 \overline{B}^0 $, $ZZ B^0 \overline{B}^0 $.
For instance, the $VVB^0 \overline{B}^0 $  couplings are
\begin{eqnarray}
L=-B_{\nu}^{(0)+} B^{(0) \nu} \left( \frac{1}{2} 
g^2 W_{\mu}^-W^{+ \mu} +
 \frac{g^2}{4 cos^2 \theta_w } Z_{\mu} Z^{\mu} 
\right)+ \nonumber \\
B_{\mu}^{(0)+} B^{(0) \nu} 
\left( \frac{1}{2} g^2 W_{\nu}^-W^{+ \mu} +
 \frac{g^2}{4 cos^2 \theta_w } Z_{\nu} Z^{\mu} \right) .
\end{eqnarray}

The processes mentioned above provide vastly different 
lower bounds on $ m_{+,0} $.
For instance, $ B^0 Z \rightarrow B^0 Z $
$ ( \overline{B}^0 Z \rightarrow \overline{B}^0 Z) $
leads to the s-wave amplitude
\begin{equation}
a_0= \left( \frac{e}{sin 2 \Theta_w} \right)^2 
\frac{m_Z^2}{32 \pi m_0^2} + {\cal O}(\frac{1}{s})
\end{equation}	
giving $ m_0 \geq 3$ GeV. 
Similarly, weak bounds emerge from elastic B--V scatterings.
For $ VV \rightarrow BB $ processes the bounds are 
better and, for instance,  $ ZZ \rightarrow B^0 \overline{B}^0 $
provides $ m_0 \geq 256$ GeV  at $ \Lambda = 1 $ TeV.
This is due to the fact that this time the 
contact and $B^0$--exchange diagrams are summed up
to an $s^2$ dependent amplitude.

The strongest lower bounds (200--400 GeV) are coming from B--B 
scatterings. Here the dominant contributions are derived 
from contact graphs. For example, in case of
$ B^0B^0 \rightarrow B^0B^0 $, the 
contribution of the Z--exchange graph to the 
lower bound of 317 GeV $ (\Lambda= 1 $ TeV ) is 4 GeV.

One finds the best bounds in  $ B^+B^- \rightarrow B^+B^- $
leading to the s--wave amplitude
\begin{equation}
a_0=- \frac{3}{16 \sqrt{10}} G_F \left( 
\frac{s^2}{2m_0m_+} -\left( \frac{m_0}{m_+} +
\frac{m_+}{m_0} \right) s 
+\frac{1}{2} m_0 m_+ \left( \frac{m_0}{m_+} +
\frac{m_+}{m_0} \right)^2 \right) .
\end{equation}
Applying the requirement of unitarity at
the maximum possible energy $ \Lambda $ , we get

\begin{eqnarray}
\Lambda=1 \hbox{TeV} &:& \quad m_0 \geq 410 \hbox{GeV}, 
\quad m_+ \geq 369 \hbox{GeV} 
\nonumber \\
\Lambda=1.5 \hbox{TeV} &:&  \quad m_0 \geq 741 \hbox{GeV}, 
\quad m_+ \geq 667 \hbox{GeV} \\
\Lambda=2 \hbox{TeV} &:&  \quad m_0 \geq 1091 \hbox{GeV}, 
\quad m_+ \geq 980\hbox{GeV}. 
\nonumber 
\end{eqnarray}

It follows that in this approximation the momentum 
scale cannot reach 2 TeV and the B bosons are heavy
particles. The bounds from $ B^+B^+ \rightarrow B^+B^+ $  are 
very close to (8) and they are in turn 
$ m_+ \geq $ 332 GeV, 615 GeV, 960 GeV. The above 
bounds imposed by the unitarity are similar to those
obtained from the S parameter:
$ m_0 \geq$ 400--550 GeV, $ m_+ \geq $ 200--350 GeV 
at $ \Lambda =$  {1 TeV} [3].

In conclusion, the vector condensate model cannot 
be renormalized perturbatively, its scattering 
amplitudes contain polynomials in s, so that 
partial--wave unitarity provides a maximum energy.
In tree--graph approximation this is 
$ \Lambda \simeq$ 2 TeV. 
A rough interpretation of the condensate
parameter d in (3) with a $ B^0 $--propagator yields
$ \Lambda \leq $2.6 TeV.

At the same time, the B--particles must be heavy 
(see (8)) and B--masses
cannot be far from $ \Lambda $.
Indeed, for $ \Lambda >> m_{+,0} $ the S parameter
becomes too large,
while the unitarity argument provides low masses
and $ \Lambda $  below $ \Lambda=${1 TeV}.

\bigskip
This work is supported in part by OTKA I/7, No. 16248.

\end{document}